%% file: ms.tex
\documentclass[a4paper,fleqn,usenatbib,useAMS,times]{mnras}
\usepackage{graphicx}	
\usepackage{amsmath}	
\usepackage{amssymb}	
\usepackage{multicol}   
\usepackage{bm}		
\usepackage{pdflscape}	
\usepackage[T1]{fontenc}
\usepackage{ae,aecompl}
\usepackage{newtxtext,newtxmath} 
\usepackage{fancyhdr}  
\usepackage{supertabular}
\usepackage{longtable}
\usepackage{url}
\usepackage{dcolumn}
\usepackage{morefloats}	 
\usepackage{rotating,caption}  	 
\usepackage{tikz}    	 
\usepackage{adjustbox}
\usepackage{blindtext} 
\usepackage{natbib} 
\usepackage{threeparttable} 
  
\newcommand{\lya}{Ly$\alpha$}

\newcommand{\HI}{\mbox{H\,{\sc i}}}

\newcommand{\CIV}{\mbox{C\,{\sc iv}}}

\newcommand{\CIII}{\mbox{C\,{\sc iii}}}

\newcommand{\zlae}{$z_{\rm peak}$} 
\newcommand{\voff}{$V_{\rm offset}$}

\newcommand{\be}{\begin{equation}}
\newcommand{\en}{\end{equation}}

\def\kms{km~s$^{-1}$}

\def\Msun{$\rm M_{\odot}$}  
\def\SFR{$\rm M_{\odot}~yr^{-1}$}   
\def\fuv{$f_{\rm UV}$}  
\def\luv{$L_{\rm UV}$}  
\title[Calibrating \lya\ redshifts]{    
{\LARGE MUSEQuBES: Calibrating the redshifts of Ly$\alpha$ emitters using stacked circumgalactic medium absorption profiles}}       
\author[S. Muzahid et al.]
{
\parbox{\textwidth}{
Sowgat Muzahid$^{1, 2}$\thanks{E-mail: sowgat@strw.leidenuniv.nl},     
Joop Schaye$^{2}$,  
Raffaella Anna Marino$^{3}$, 
Sebastiano Cantalupo$^{3}$, 
Jarle Brinchmann$^{2, 4}$, 
Thierry Contini$^{5}$, 
Martin Wendt$^{1, 6}$,  
Lutz Wisotzki$^{1}$, 
Johannes Zabl$^{7}$,      
Nicolas Bouch{\'e}$^{7}$,   
Mohammad Akhlaghi$^{8, 9}$,    
Hsiao-Wen Chen$^{10}$, 
Ad{\'e}la$\ddot{\rm i}$de Claeyssens$^{7}$,   
Sean Johnson$^{11, 12}$\thanks{Carnegie--Princeton fellow}, 
Floriane Leclercq$^{7, 13}$,  
Michael Maseda$^{2}$, 
Jorryt Matthee$^{3}$, 
Johan Richard$^{7}$,  
Tanya Urrutia$^{1}$,   
and 
Anne Verhamme$^{13}$  
} 
\vspace*{8pt} \\    
$^{1}$ Leibniz-Institut f$\ddot{u}$r Astrophysik Potsdam (AIP), An der Sternwarte 16, D-14482 Potsdam, Germany \\  
$^{2}$ Leiden Observatory, Leiden University, PO Box 9513, NL-2300 RA Leiden, the Netherlands  \\ 
$^{3}$ Department of Physics, ETH Z$\ddot{u}$rich, Wolfgang-Pauli-Strasse 27, 8093 Z$\ddot{u}$rich, Switzerland \\
$^{4}$ Instituto de Astrof{\'\i}sica e Ci{\^e}ncias do Espa{\c{c}}o, Universidade do Porto, CAUP, Rua das Estrelas, PT4150-762 Porto, Portugal \\  
$^{5}$ Institut de Recherche en Astrophysique et Plan$\acute{e}$tologie (IRAP), Universit$\acute{e}$ de Toulouse, CNRS, UPS, F-31400 Toulouse, France \\ 
$^{6}$ Institut f$\ddot{u}$r Physik und Astronomie, Universit$\ddot{a}$t Potsdam, Karl-Liebknecht-Str 24/25, D-14476 Golm, Germany \\   
$^{7}$ Univ Lyon, Univ Lyon1, Ens de Lyon, CNRS, Centre de Recherche Astrophysique de Lyon UMR5574, F-69230, Saint-Genis-Laval, France \\ 
$^{8}$ Instituto de Astrof{\'\i}sica de Canarias, C/ V{\'i}a L{\^a}ctea, 38200 La Laguna, Tenerife, Spain \\  
$^{9}$ Facultad de F{\'\i}sica, Universidad de La Laguna, Avda. Astrof{\'\i}sico Fco. S{\^a}nchez s/n, 38200 La Laguna, Tenerife, Spain \\ 
$^{10}$ Department of Astronomy \& Astrophysics, The University of Chicago, Chicago, IL 60637, USA  \\ 
$^{11}$ Department of Astrophysical Science, 4 Ivy Lane, Princeton University, Princeton, NJ 08644, USA \\ 
$^{12}$ The Observatories of the Carnegie Institution for Science, 813 Santa Barbara Street, Pasadena, CA 91101, USA \\ 
$^{13}$ Observatoire de Gen$\acute{e}$ve, Universit$\acute{e}$ de Gen$\acute{e}$ve, 51 Ch. des Maillettes, CH-1290 Versoix, Switzerland \\ 
}   
\date{Accepted. Received; in original form} 
\pubyear{2019}
\begin{document}
\label{firstpage}
\pagerange{\pageref{firstpage}--\pageref{lastpage}}   
\maketitle   
\begin {abstract} 

\noindent 
Lyman$-\alpha$ (\lya) emission lines are typically found to be redshifted with respect to the systemic redshifts of galaxies, likely due to resonant scattering of \lya\ photons. Here we measure the average velocity offset for a sample of 96 $z\approx3.3$ \lya\ emitters (LAEs) with a median \lya\ flux (luminosity) of $\approx 10^{-17}~\rm erg~cm^{-2}~s^{-1}$ ($\approx10^{42}~\rm erg~s^{-1}$) and a median star formation rate (SFR) of $\approx1.3$~\SFR\ (not corrected for possible dust extinction), detected by the Multi-Unit Spectroscopic Explorer as part of our MUSEQuBES circumgalactic medium (CGM) survey. By postulating that the stacked CGM absorption profiles of these LAEs, probed by 8 background quasars, must be centered on the systemic redshift, we measure an average velocity offset, \voff~$=$~171$\pm$8~\kms, between the \lya\ emission peak and the systemic redshift. The observed \voff\ is lower by factors of $\approx1.4$ and $\approx2.6$ compared to the velocity offsets measured for narrow-band selected LAEs and Lyman break galaxies, respectively, which probe galaxies with higher masses and SFRs. Consistent with earlier studies based on direct measurements for individual objects, we find that the \voff\ is correlated with the full width at half-maximum of the red peak of the \lya\ line, and anti-correlated with the rest-frame equivalent width. Moreover, we find that \voff\ is correlated with SFR with a sub-linear scaling relation, $V_{\rm offset}\propto \rm SFR^{0.16\pm0.03}$. Adopting the mass scaling for main sequence galaxies, such a relation suggests that \voff\ scales with the circular velocity of the dark matter halos hosting the LAEs.  
\end {abstract}


\section{Introduction} 
\label{sec:intro}

Lyman-$\alpha$ (\lya) emitters (LAEs) are galaxies that are identified through the \lya\ line of neutral hydrogen ($\lambda1215.67$~\AA). Owing to the high cosmic abundance of hydrogen and the large oscillator strength of the $2p \longrightarrow 1s$ transition, \lya\ emission has been recognized as an excellent tool to identify galaxies using a variety of techniques, including narrow-band (NB) and medium-band surveys \citep[e.g.,][]{Malhotra02,Gronwall07,Sobral18,Shibuya18}, integral-field-spectroscopy (IFS) surveys \citep[e.g.,][]{Wisotzki16,Inami17,Leclercq17,Urrutia19}, multi-object spectroscopy \citep[e.g.,][]{Cassata11}, and long-slit spectroscopy \citep[e.g.,][]{Rauch08,Rauch16}. LAEs detected via different techniques can probe a diverse galaxy population, however, there is a growing consensus that the majority of LAEs are typically low-mass, star-forming galaxies \citep[e.g.,][]{Gawiser07,Hagen16,Hao18}.        

Though \lya\ is an excellent tool to detect galaxies, particularly at high redshift ($z>2$), interpreting \lya\ emission spectra is challenging because of resonant scattering and susceptibility to dust extinction \citep[e.g.,][]{Hayes15}. The \lya\ spectrum emerging from a uniform spherical, static gas cloud with a central \lya\ emitting source appears as symmetric double peaked emission with a peak separation that increases with increasing line-center optical depth \citep[e.g.,][]{Neufeld90,Zheng02,Cantalupo05,Verhamme06,Dijkstra06a}. Any bulk motion of the gas with respect to the central source, however, makes the peaks asymmetric. For example, outflowing (infalling) gas would enhance the red\footnote{The lower energy (higher wavelength) peak.} (blue) peak and suppress the blue (red) peak \citep[e.g.,][]{Laursen09}. In fact, the signature of outflowing gas (i.e., a dominant asymmetric red peak and an occasional weaker blue ``bump'') is ubiquitous in the spectra of high-$z$ LAEs \citep[e.g.,][]{Gronke17}. Composite spectra of high-$z$ LAEs indeed show signatures of metal enriched outflows with outflow velocity increasing with continuum luminosity \citep[]{Trainor15}. Owing to resonant scattering, the \lya\ emission line does not trace the systemic redshift. In fact, observations have shown that \lya\ redshifts are, on average, shifted by $\approx+230$~\kms\ \citep[for LAEs; e.g.,][]{Shibuya14} to $\approx+440$~\kms \citep[for Lyman break galaxies (LBGs); e.g.,][]{Steidel10}. The \lya\ redshifts should, thus, be taken with caution in the absence of non-resonant rest-frame ultraviolet (UV)/optical stellar absorption and/or nebular emission lines which provide the most accurate galaxy redshifts. 

Recently, \citet{Verhamme18} suggested two empirical relations to recover the systemic redshift of galaxies from their \lya\ line profile using the observed correlations between (i) the velocity offset (measured from non-resonant UV/optical lines) and the full width at half-maximum (FWHM) of the red peak of the \lya\ line; (ii) the velocity offset and the velocity separation between the red peak and the blue bump. \citet{Erb14} reported $>3\sigma$ correlations between velocity offset and $R-$band magnitude, $M_{UV}$, and the velocity dispersion of nebular emission lines for a sample of 36 LAEs at $z\approx2-3$. In addition, a strong anti-correlation ($>7\sigma$) was found between velocity offset and the \lya\ equivalent width ($\rm EW_{0}$). Such empirical relationships are valuable for understanding the physics of the \lya\ emitting galaxies, and provide indirect means to obtain the systemic redshifts. Finding and confirming such empirical relations and observational trends using complementary techniques is thus important.  

Obtaining accurate systemic redshifts is particularly important for studying the circumgalactic medium (CGM) of galaxies using background quasars, since the association of galaxies with their CGM absorption lines, seen in the quasar spectrum, is based on velocity coincidence. CGM studies in the literature typically adopt a velocity window of $\pm500$~\kms\ around the galaxy redshift to search for associated CGM absorption. It is thus essential to know the galaxy redshifts with an accuracy of $\Delta z/(1+z) \approx 10^{-3}$ or better. Using guaranteed time observations with the Multi-Unit Spectroscopic Explorer \citep[MUSE;][]{Bacon10}, we conducted the MUSEQuBES (MUSE Quasar-field Blind Emitters Survey) survey-- a blind search for LAEs in $1'\times1'$ fields centered on 8 bright $z\approx3.6-3.8$ quasars (see Table~\ref{tab:observation}). This is the first systematic survey of the CGM of LAEs in absorption (\color{blue} Muzahid et al., \color{black}in preparation). Since, we generally do not have access to stellar absorption and/or non-resonant nebular emission lines for the LAEs in our sample, we must make use of the \lya\ redshifts (\zlae; determined from the peak of the \lya\ line). Here we adopt the approach proposed by \citet[]{Rakic11} to calibrate the \lya\ redshifts in a statistical manner using mean/median stacked CGM\footnote{Note that, in most cases, the impact parameters of the LAEs in our sample are larger than the inferred virial radii (the latter are a few tens of kpc).} absorption (\HI~\lya) profiles, by requiring that the average CGM absorption profiles must be centered on the systemic velocity since the LAEs are randomly oriented with respect to the background quasar. \citet[]{Rakic11} applied this technique to a large sample of $z\approx2.3$ LBGs, finding velocity offsets that agreed with the direct measurements from non-resonant nebular lines available for a subset of their sample.

This paper is organized as follows: In Section~\ref{sec:observations} we briefly describe the observations and data reduction procedures. In Section~\ref{sec:sample} we summarize the properties of our LAE sample. Section~\ref{sec:results} presents the main results, followed by a discussion in Section~\ref{sec:diss}. Section~\ref{sec:con} concludes the paper. Throughout this study, we adopt a flat $\Lambda$CDM cosmology with $H_0=70$~\kms~Mpc$^{-1}$, $\Omega_{\rm M}=0.3$ and $\Omega_{\Lambda}=0.7$. All distances given are in physical units.  

\begin{table}  
\begin{threeparttable}[b] 
\caption{The data sample}   
\begin{tabular}{lcccrr}   
\hline  
Quasar Field   &  $\rm RA_{QSO}$  &  $\rm Dec_{QSO}$  &  $z_{\rm QSO}$   &   $t_{\rm exp}$  &    $\rm N_{LAE}$   \\ 
(1)            &  (2)             &  (3)              &  (4)             &  (5)                    &  (6)               \\  
\hline  
Q1422$+$23     &  14:24:38.1 & $+$22:56:01  &	3.620	&  4    &    8	 \\ 	 	
Q0055$-$269    &  00:57:58.1 & $-$26:43:14  &	3.655	& 10	&   12	 \\     	
Q1317$-$0507   &  13:20:30.0 & $-$05:23:35  &	3.700	& 10	&   22	 \\     
Q1621$-$0042   &  16:21:16.9 & $-$00:42:50  &	3.709	&  9	&   12	 \\     
QB2000$-$330   &  20:03:24.0 & $-$32:51:44  &	3.773	& 10	&   14 	 \\     
PKS1937$-$101  &  19:39:57.3 & $-$10:02:41  &	3.787	&  3	&    2	 \\    	 
J0124$+$0044   &  01:24:03.0 & $+$00:44:32  &	3.834	&  2	&    4	 \\      
BRI1108$-$07   &  11:11:13.6 & $-$08:04:02  &	3.922	&  2	&   22	 \\       
\hline  
\end{tabular} 
\label{tab:observation}    
Notes-- (1) Name of the quasar field; (2) Right Ascension (J2000), (3) Declination (J2000), and (4) Redshift of the quasar; (5) MUSE exposure time of the field in hour; (6) Number of detected LAEs in the redshift range of interest.    
\end{threeparttable}  
\end{table} 

\section{Observations \& Data Reduction}       
\label{sec:observations}  

Our MUSEQuBES survey utilizes $\approx50$h of MUSE GTO observations in the wide field mode centered on 8 high-$z$ quasars that have high quality ($\rm S/N>50$ per pixel) optical spectra obtained with the Very Large Telescope Ultraviolet and Visual Echelle Spectrograph (VLT/UVES) and/or Keck HIgh Resolution Echelle Spectrometer (Keck/HIRES). The details of the quasar fields are given in Table~\ref{tab:observation}. The MUSE and UVES/HIRES observations and data reduction procedures will be presented in \color{blue}Muzahid et al., \color{black} (in preparation). The MUSE data for four of the eight quasar fields (Q0055$-$269, Q1317$-$0507, Q1621$-$0042, and Q2000$-$330) were reduced using the standard MUSE pipeline {\tt v1.6} \citep[]{Weilbacher20} and post-processed with the tools in the CubExtractor package \citep[{\tt CubEx v1.6}; Cantalupo in preparation; see][for a description]{Cantalupo19} to improve flat-fielding and sky subtraction as presented in \citet[]{Marino18}. The remaining fields were reduced using the MUSE pipeline software ({\tt v2.4}) and post-processed with {\tt CubEx v1.6} following the same procedures. 

The optical spectra of the quasars were obtained primarily using VLT/UVES with resolution, $R\approx45000$. The final coadded and continuum normalized spectra were downloaded from the SQUAD database \citep[]{Murphy19} for all but Q1422$+$23. The spectrum of the quasar Q1422$+$23 was reduced using the Common Pipeline Language ({\tt CPL v6.3}) of the UVES pipeline. After the standard reduction, the custom software UVES Popler\footnote{https://doi.org/10.5281/zenodo.44765} was used to combine the extracted echelle orders into single 1D spectra. The coadded spectrum was continuum normalized by low-order spline interpolation to the absorption line free regions determined by iterative sigma-clipping. Keck/HIRES data are available for four quasars: BRI1108--07, PKS1937--101, QB2000--330, and Q1422+23. We used the HIRES spectra of PKS1937--101 and Q1422+23 from the KODIAQ data release \citep[]{OMeara15} to fill in the gaps in the UVES spectra. We combined the continuum normalized UVES and HIRES spectra using inverse-variance weighting. Air-to-vacuum conversion was done for both the MUSE and UVES spectra before performing any measurements. 

For the CGM spectral stacking analysis we have first used the pixel optical depth method \citep[][]{Cowie98,Aguirre02,Schaye03,Turner14} using the python module {\tt PODPy} developed by \citet{Turner14}. {\tt PODPy} iteratively examines whether the optical depth of a given pixel in a quasar spectrum is consistent with being the transition of interest (e.g., \lya, \CIV). {\tt PODPy} corrects for contamination by flagging the pixels whose optical depths are not consistent with the expectation. For multiplets, it uses all available transitions (up to Ly$\eta$ for \HI) leading to a larger dynamic range in the ``recovered'' optical depth. We refer the reader to Appendix~A of \citet{Turner14} for more details. The optical depth recovered by {\tt PODPy} was then converted to flux before stacking the quasar spectra.

\begin{figure}
\centerline{\vbox{
\centerline{\hbox{ 
\includegraphics[width=0.50\textwidth,angle=00]{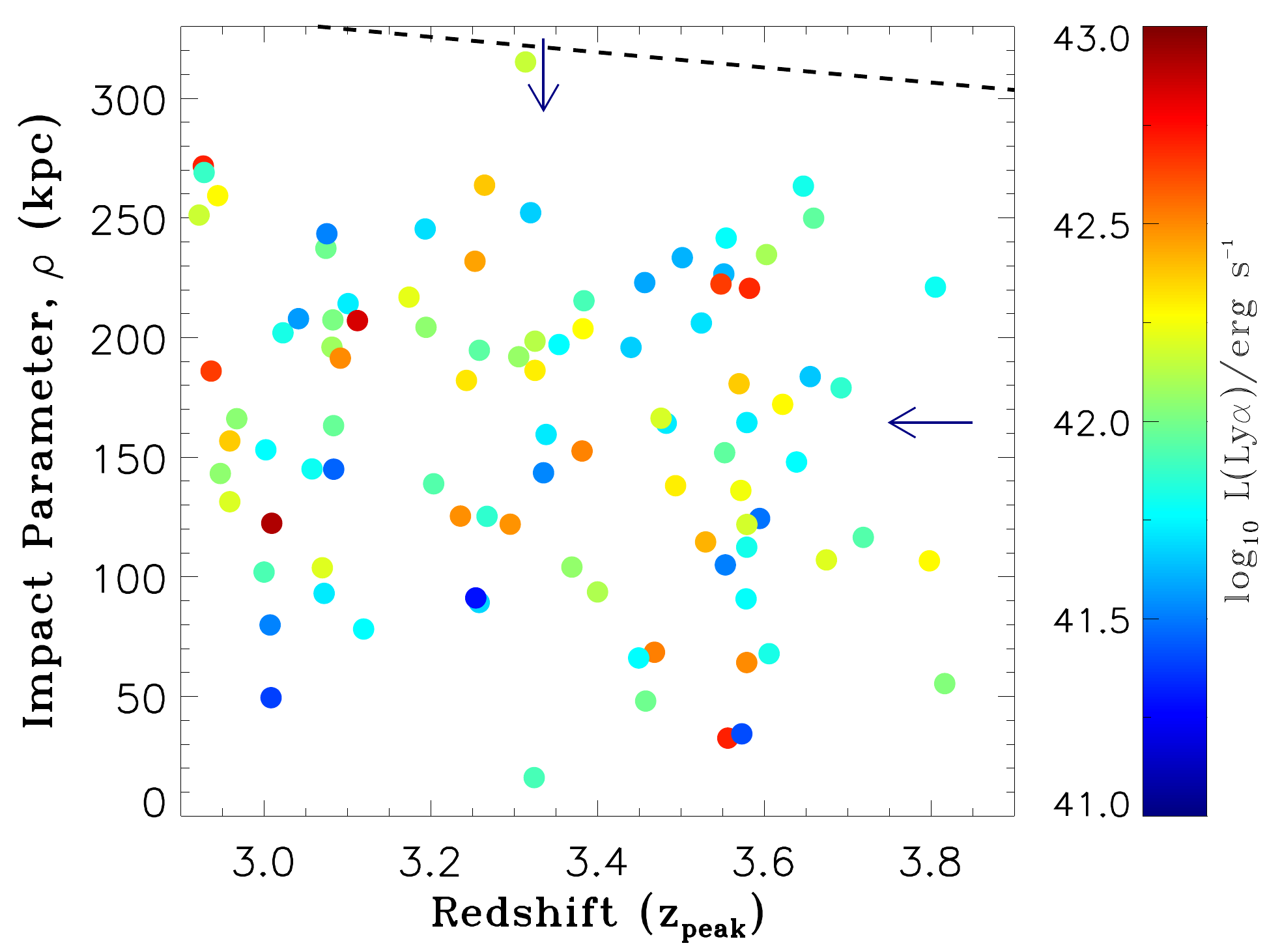} 
}}
}}   
\caption{Scatter plot of impact parameter versus redshift of the 96 LAEs. The data points are color coded by the \lya\ luminosity. The median impact parameter and median redshift are indicated by the arrows. The maximum impact parameter allowed by the MUSE FoV is shown by the dashed line on the top.}        
\label{fig:sample}  
\end{figure} 

\section{The LAE sample}       
\label{sec:sample}      

We used {\tt CubEx v1.6} \citep[][]{Cantalupo19} for automatic extraction of emission line sources in the MUSE datacubes. {\tt CubEx} uses a 3D extension of the connected-component labeling algorithm. The source extraction and classification procedure will be presented in detail in \color{blue}Muzahid et al., \color{black}(in preparation). Briefly, after spatially smoothing (by 2 pixels radius) the datacubes and the corresponding variances at each wavelength layer, we require three conditions to be satisfied for a detection: (i) $\rm S/N$ per voxel $ > 4.0$, (ii) number of connected voxels $N_{\rm vox} > 40$, and (iii) a spectral $\rm S/N > 4.5$ measured on the 1D \lya\ emission line spectrum. All the extracted objects are then visually inspected (both 1D spectra, extracted using the segmentation  maps produced  by {\tt CubEx}, and pseudo-NB images around the emission features) and classified by two members of the team (SM and RAM) independently. 

A total of 96 LAEs have been detected in the 8 MUSE cubes. The impact parameters ($\rho$) of the LAEs from the quasar sight lines are plotted against the \lya\ peak redshifts (\zlae) in Fig.~\ref{fig:sample}. The redshifts of the LAEs are determined directly from the peak of the emission lines in the 1D spectra without any modelling. We ensured that \zlae\ is not affected by noise-spikes by visually inspecting the spectra. In case of double peaked emission we used the red peak for the \zlae\ measurement. The minimum \zlae\ ($\approx$2.9) is determined by the lowest wavelength covered by MUSE. Note that we did not use the first 8--10 wavelength layers ($\approx10$~\AA) in our search, in order to avoid a large number of spurious detections at the very edge of the spectrum. The maximum \zlae\ is determined by the quasar redshift ($z_{\rm QSO}$). In order to exclude the quasars' proximity regions \citep[see e.g.,][]{Muzahid13}, we did not use the 3000~\kms\ bluewards of the $z_{\rm QSO}$. The LAEs in our sample span a redshift range of 2.92--3.82 with a median \zlae\ of 3.33. The maximum and minimum $\rho$ values are determined by the MUSE field-of-view and the quasars' point spread functions (PSFs), respectively. The $\rho$ values span 16--315~kpc with a median of 165~kpc. 

\begin{figure*}
\centerline{\vbox{
\centerline{\hbox{ 
\includegraphics[width=0.85\textwidth,angle=00]{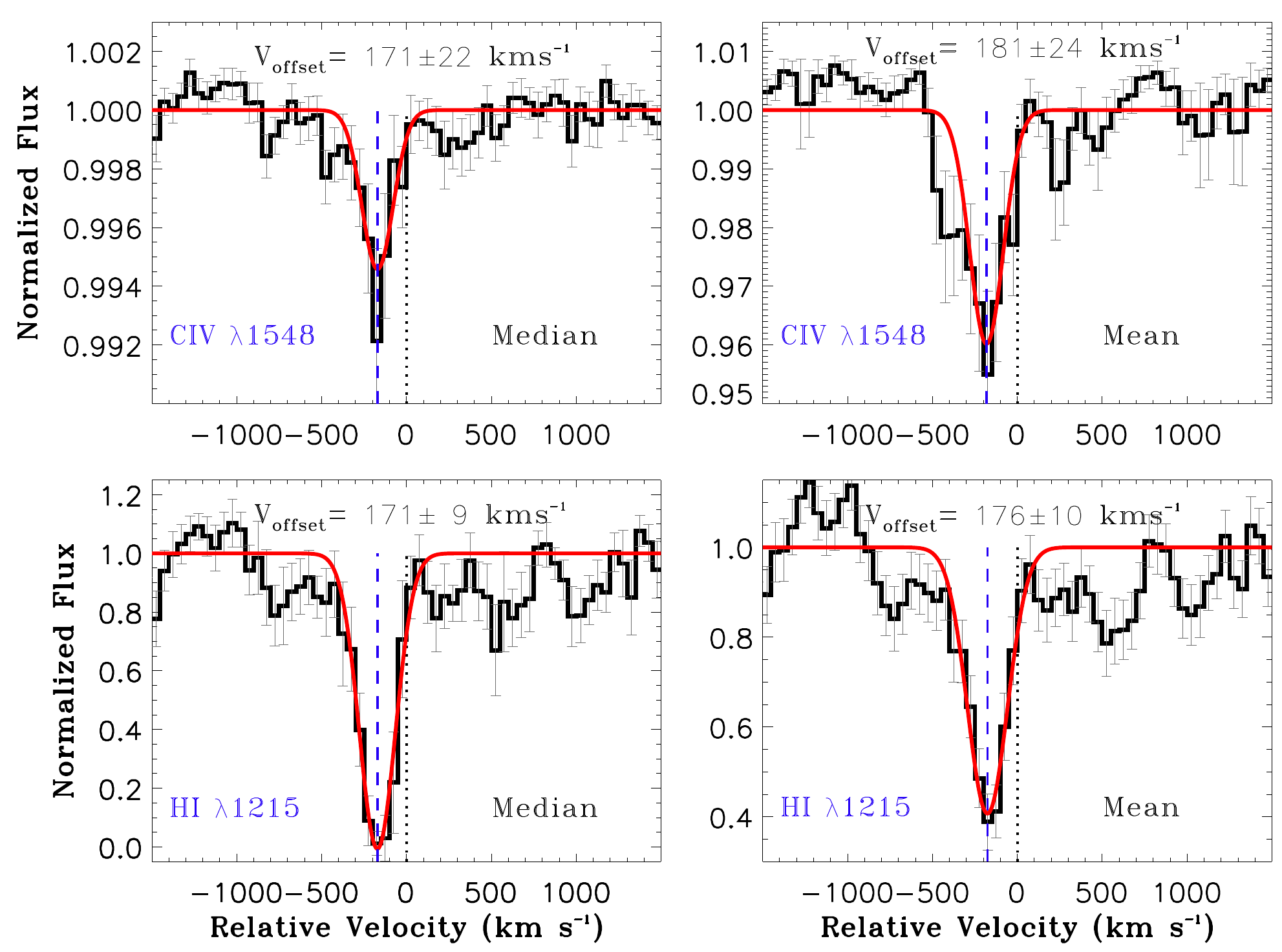} 
}}
}}   
\caption{Median (left) and mean (right) stacked CGM absorption profiles of \HI~$\lambda1215$ (bottom) and \CIV~$\lambda1548$ (top). The zero velocity ($V_{\rm peak}$) is defined by \zlae, the redshift of peak \lya\ emission. The profiles are normalized to the pseudo-continuum estimated far away from zero velocity. The $1\sigma$ errors are calculated from 1000 bootstrap realizations of the LAE sample. The best-fitting Gaussian profiles are shown by the smooth red curves. The centroids of the Gaussians ($V_{\rm CGM}$), marked by the blue vertical dashed lines, provide the velocity offset, \voff~$\equiv (V_{\rm peak} - V_{\rm CGM})$. \voff\ measured for the different stacked profiles are indicated in the corresponding panels. The weighted average of the \voff\ values is 171$\pm$8~\kms\ (177$\pm$9~\kms) for the median (mean) stacked profiles. 
} 
\label{fig:profiles}   
\end{figure*} 

The data points in Fig.\ref{fig:sample} are color coded by the \lya\ luminosity, $L$(\lya), calculated from the Galactic extinction corrected  line flux, $f$(\lya)\footnote{We used the $\rm E(B-V)$ values from \citet{Schlafly11} and the \citet{Fitzpatrick99} extinction curve to de-redden the fluxes.}. The $f$(\lya) values are measured from pseudo-NB images using the curve-of-growth method following \citet{Marino18}. The $f$(\lya) values are found to be in the range $10^{-17.7} - 10^{-16.0}$~$\rm erg~cm^{-2}~s^{-1}$ with a median value of $10^{-17.0}~\rm erg~cm^{-2}~s^{-1}$. The $L$(\lya) spans $10^{41.3} - 10^{42.9}~\rm erg~s^{-1}$ with a median value of $10^{42.0}~\rm erg~s^{-1}$. Following \citet{Verhamme18}, the FWHM (of the red peak for the handful of double peaked profiles) is calculated directly from the 1D spectrum, without any modelling and without correcting for instrumental broadening, as the velocity width of the \lya\ emission line with flux above half of the maximum flux value. The FWHM values span the range 120--528~\kms\ with a median value of 240~\kms. Here we note that 10 LAEs show FWHM lower than the MUSE spectral resolution of $\approx166$~\kms\ at the median \lya\ wavelength of our sample.

The UV continuum fluxes, \fuv, and the associated errors are derived by integrating the 1D spectra, extracted from the original cubes (not continuum subtracted) and the corresponding variance cubes using the same segmentation maps used to obtain the 1D \lya\ emission spectra. We chose a wavelength range of rest-frame 1410--1640~\AA, the same as the wavelength range covering the FWHM of the $GALEX$ far-UV transmission curve. No \fuv\ are calculated for the 15 LAEs that are contaminated by low-$z$ continuum sources. About 48\% (39/81) of the remaining LAEs are detected in UV continuum emission with $>5\sigma$ significance. For the 39 continuum detected objects, \fuv\ values (corrected for Galactic extinction) are in the range $10^{-17.0}-10^{-15.6}~\rm erg~cm^{-2}~s^{-1}$ with a median of $10^{-16.4}~\rm erg~cm^{-2}~s^{-1}$. For the remaining 42 LAEs for which we could place meaningful $5\sigma$ upper limits, the \fuv\ values were found to be lower than $10^{-16.4}~\rm erg~cm^{-2}~s^{-1}$. The UV continuum luminosity, \luv, ranges from $10^{42.1}-10^{43.4}~\rm erg~s^{-1}$ for the continuum detected objects (median $10^{42.7}~\rm erg~s^{-1}$ ). For the non-detections, the upper limits on \luv\ are in the range of $10^{41.9}-10^{42.7}~\rm erg~s^{-1}$.

The dust-uncorrected SFRs are calculated from the measured \luv\ values using the local calibration relation of \citet{Kennicutt98} corrected to the \citet{Chabrier03} initial mass function \citep[IMF; see][]{Madau14}. The SFRs for the continuum detected LAEs span $0.3-7.1$~\SFR\ with a median SFR of $1.3$~\SFR. For the continuum un-detected LAEs, the SFRs are $<1.5$~\SFR. The rest-frame equivalent width of the \lya\ emission ($\rm EW_{0}$) is obtained by dividing the \lya\ line flux by the continuum flux density and then divided by $(1+z_{\rm peak})$. The continuum flux density is estimated from the extrapolation of the measured continuum at rest-frame 1500~\AA\ assuming a UV continuum slope ($\beta_{\rm UV}$) of $-2.0$ \citep{Bouwens14}. The continuum detected objects have $\rm EW_0$ in the range 9--113~\AA\ with a median $\rm EW_{0}$ of 48~\AA.

In our redshift range of interest ($z\approx$~3--4), the presence of the non-resonant \CIII]~$\lambda\lambda$1907,1909 doublet in the MUSE spectra is an excellent means to obtain the systemic redshift. We detect the \CIII]~$\lambda\lambda$1907,1909 doublet for only 3 LAEs, one of them being tentative. Such a low detection rate of the \CIII] line is consistent with the recent results of \citet{Maseda17}.

\input{table}


%
\begin{figure*}
\centerline{\vbox{
\centerline{\hbox{ 
\includegraphics[trim=10 190 10 240,clip,width=0.70\textwidth,angle=00]{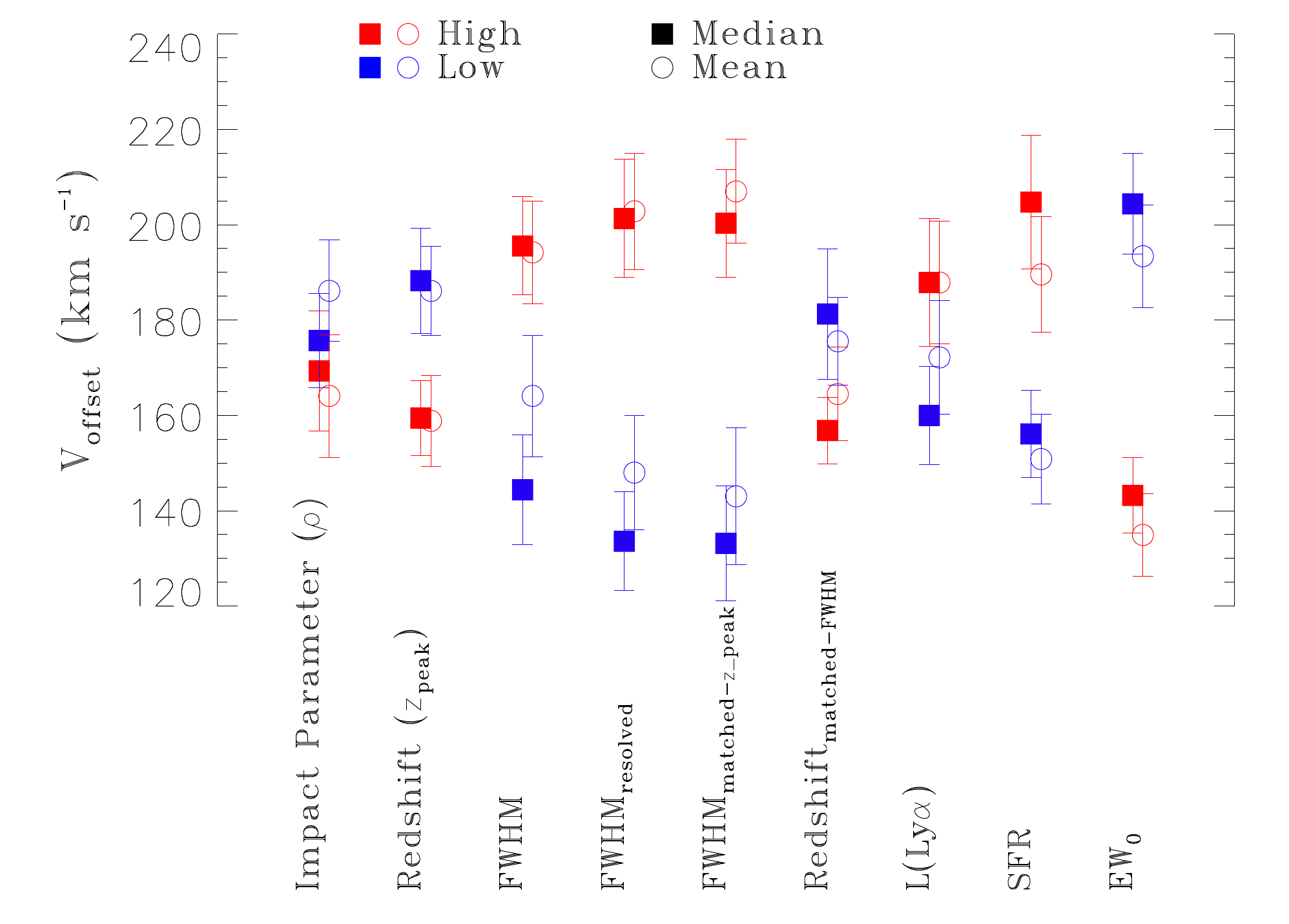} 
}}
}}   
\caption{The inverse variance weighted average of the velocity offsets measured from the \HI\ and \CIV\ absorption profiles for the different sub-samples listed in Table~\ref{tab:offset_summary}. The velocity offsets measured from the median and mean stacked profiles are indicated by the filled squares and open circles respectively. The red and blue points correspond to the ``high'' and ``low'' sub-samples (Table~\ref{tab:offset_summary}), respectively. \voff\ shows significant trends with FWHM, SFR, and EW$_{0}$.}          
\label{fig:offset_summary}     
\end{figure*} 


\section{Results}     
\label{sec:results}

The median and mean stacked absorption profiles, normalized to the pseudo-continua estimated at large velocities, of \HI\ \lya\ and \CIV, arising from the CGM of the LAEs, are shown in Fig.~\ref{fig:profiles}. For each transition (\HI~$\lambda1215$ or \CIV~$\lambda1548$), we have selected the part of the quasar spectrum covering a velocity range of $-3000$ to $+3000$~\kms\ with respect to the \zlae\ for a given LAE. The mean and median fluxes for the full sample are then calculated from the {\tt PODPy} ``recovered'' pixel optical depths in bins of 50~\kms. We note here that our main conclusions remain valid even if we use the original quasar spectra for stacking instead of using {\tt PODPy} recovered spectra. However, in that case the stacked profiles become noisier, particularly when we split the sample into different sub-samples. We thus chose to use the {\tt PODPy} recovered spectra. The stacked profiles for the full sample using the original spectra are shown in Fig.~\ref{fig:prof_comp}.

Fig.~\ref{fig:profiles} shows the first measurements of the CGM of LAEs in absorption for a statistically meaningful sample \citep[see][for individual examples]{Diaz15,Fumagalli16,Zahedy19b,Mackenzie19,Lofthouse20}. We detect absorption signals for \HI\ and \CIV\ with $>5\sigma$ significances. Note that, none of the stacked absorption profiles are centered on the $0$~\kms\ defined by the redshift of peak \lya\ emission, \zlae. All profiles show velocity offset, \voff~$>$170~\kms. Here \voff~$= (V_{\rm peak} - V_{\rm CGM})$, where $V_{\rm CGM}$ is the velocity centroid of the CGM absorption profile, and $V_{\rm peak}$ is the velocity corresponding to \zlae. The \voff\ measured for the median (mean) stacked \HI\ profile is 171$\pm$9~\kms\ (176$\pm$10~\kms). The velocity offsets and the associated errors are determined from Gaussian fits to the stacked spectra along with the error spectra determined by bootstrapping the LAE sample.\footnote{Note that the possible covariance between the neighboring pixels are ignored in our fits. The \voff\ distributions obtained from the bootstrapped spectra suggest that such ignorance could lead to underestimations of the true uncertainties by a factor of 1.3--1.5 for the \HI\ profiles.} The median and mean stacked \CIV\ profiles show \voff\ of 171$\pm$22~\kms and 181$\pm$24~\kms, respectively. Owing to the relative weakness of the \CIV\ absorption, the estimated errors on the corresponding \voff\ measurements are larger. Nevertheless, the stacked \CIV\ profiles provide independent measurements of \voff, and are fully consistent with the \HI\ measurements. The weighted average of the \voff\ values measured from the median and mean stacked profiles are 171$\pm$8~\kms\ and 177$\pm$9~\kms, respectively. Such offsets imply that the \zlae\ values are systematically redshifted with respect to the systemic redshifts, consistent with the results from the observations of non-resonant rest-frame UV/optical nebular emission/absorption lines \citep[e.g.,][]{Steidel10,Shibuya14,Verhamme18}.

In order to investigate possible trends between \voff\ and other parameters (e.g., \zlae, $\rho$, FWHM), we generated stacked \HI\ and \CIV\ absorption profiles for several sub-samples corresponding to different parameters, as summarized in Table~\ref{tab:offset_summary}. The velocity offsets and corresponding uncertainties, determined from Gaussian fits to the stacked \HI\ and \CIV\ profiles (as in Fig.~\ref{fig:profiles}), for the different sub-samples are also listed in the table. The last two columns (columns 11 \& 12) provide the combined constraints on \voff, obtained from the inverse variance weighted average of the velocity offsets measured from the \HI\ and \CIV\ profiles, and are illustrated in Fig.~\ref{fig:offset_summary}. We will only use these weighted average \voff\ values in all further discussions.

It is evident from Fig.~\ref{fig:offset_summary} that \voff\ does not show any significant trend with $\rho$ and $L$(\lya). The difference in \voff\ between the corresponding ``high'' and ``low'' sub-samples, calculated for both the mean and median stacked profiles, has $<$2$\sigma$ significance. There is a 2.2$\sigma$ (2.2$\sigma$) difference between the \voff\ values measured from the median (mean) stacked profiles of the low-- and high--\zlae\ sub-samples. However, we note that the trend is actually driven by FWHM, owing to a $3.4\sigma$ anti-correlation between \zlae\ and FWHM (Spearman rank correlation coefficient, $r_s=-0.35$)\footnote{The anti-correlation between \zlae\ and FWHM is likely due to the fact that the MUSE resolution improves from $\approx180$~\kms\ to $\approx150$~\kms\ between $z\approx3.0$ and $3.6$.}. Indeed, the difference reduces to $<2\sigma$ when the low-- and high--\zlae\ sub-samples are matched in FWHM.

There is a 3.4$\sigma$ (1.9$\sigma$) difference in the \voff\ for the high--FWHM and low--FWHM sub-samples measured from the median (mean) stacked profiles. The difference increases for the $\rm FWHM_{resolved}$ sub-samples, in which we excluded the LAEs with FWHM smaller than the MUSE resolution, to 4.4$\sigma$ (3.3$\sigma$ for the mean stack). Since we noted an anti-correlation between \zlae\ and FWHM, it is important to investigate whether the trend between \voff\ and FWHM remains when the low-- and high--FWHM sub-samples are matched in \zlae. In fact, we do find a 4.1$\sigma$ (3.7$\sigma$ for the mean stack) difference in \voff\ between the low-- and high--FWHM sub-samples when they are matched in \zlae. In addition, a clear difference, with $>2.5\sigma$ significance, is seen in \voff\ measured for the low-- and high--SFR sub-samples, for both the mean and median stacked profiles. Finally, the strongest difference ($>4.5\sigma$) in \voff\ is seen between the low-- and high--EW$_0$ sub-samples, with higher EW$_0$ yielding a smaller velocity offset. In the next section we discuss the possible implications of these new results in the context of existing observational and theoretical studies.

\section{Discussion}    
\label{sec:diss}

\begin{figure*}
\centerline{\vbox{
\centerline{\hbox{ 
\includegraphics[width=0.50\textwidth,angle=00]{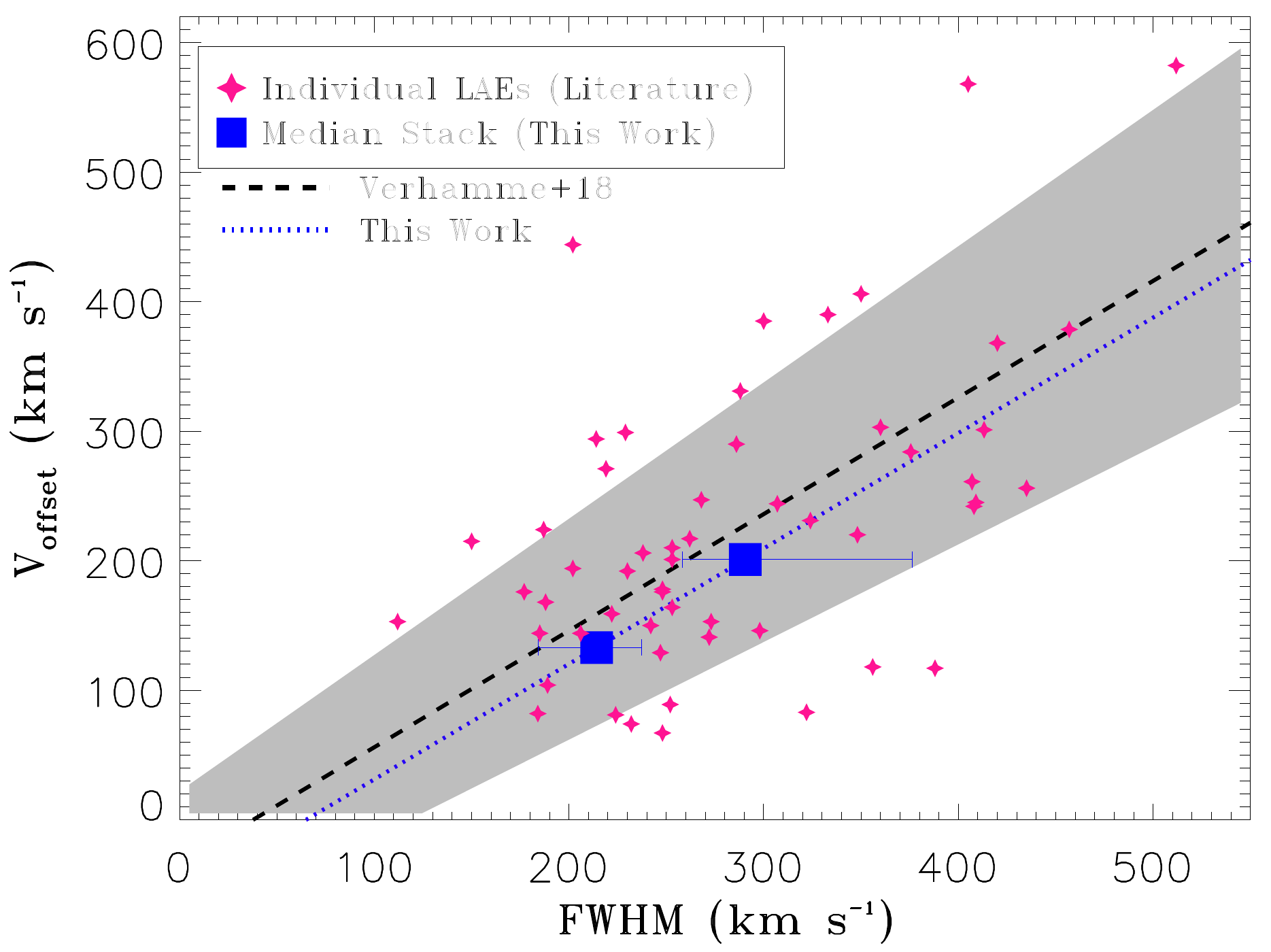} 
\includegraphics[width=0.50\textwidth,angle=00]{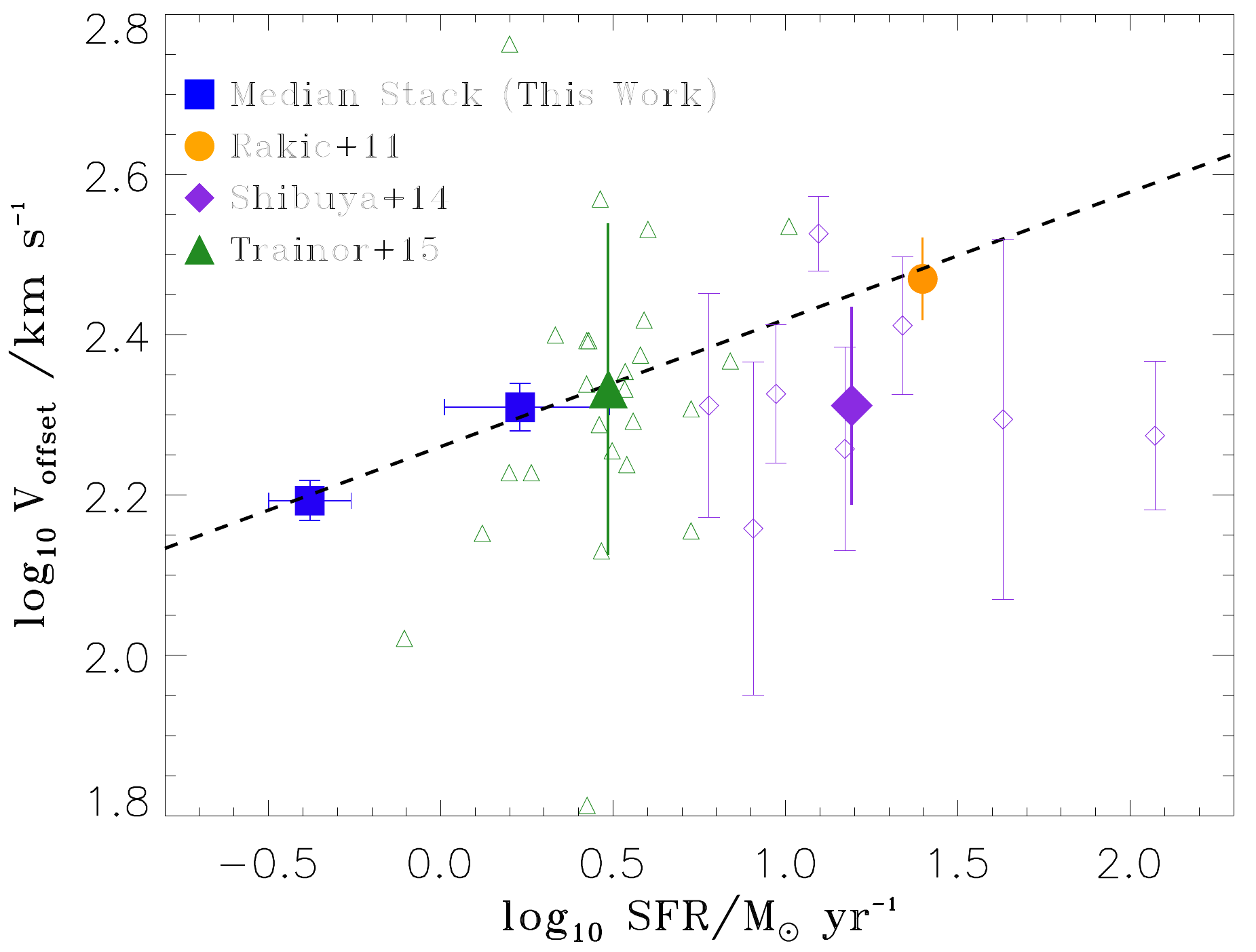} 
}}
}}   
\caption{{Left:} \voff\ as a function of FWHM of the \lya\ line. The star symbols represent LAEs from the literature for which the systemic redshifts are known (see \citet{Verhamme18} for details). The dashed line and the shaded region represent the best-fit linear relation, \voff $= 0.9(\pm0.14)\times \rm FWHM - 34(\pm60)$~\kms, for the star symbols and its $1\sigma$ range, respectively, as found by \citet{Verhamme18}. The blue squares represent our weighted average \voff\ measurements for the FWHM$_{\rm resolved}$ sub-samples as in Table~\ref{tab:offset_summary}. The dotted straight line connecting the blue squares has a slope (0.89) and an intercept ($-$58~\kms) that are fully consistent with the empirical relation of \citet{Verhamme18}.  
{Right:} Weighted average \voff, measured from the median stacked profiles, as a function of SFR for our sample (blue squares). The orange filled circle represents the measurement from \citet{Rakic11} for $z\approx2.3$ LBGs. The best-fit linear relation between $\log$~\voff\ and $\log~\rm SFR$ ($\log$~\voff~$=(0.16\pm0.03)\log \rm SFR + (2.26\pm0.02)$) for those three points is indicated by the dashed line. 
The open diamonds and the open triangles  represent data points from \citet[][]{Shibuya14} and \citet[]{Trainor15}, respectively (see text). The filled diamond and the filled triangle represent the median SFR and the median \voff\ of the corresponding samples, with the error bars indicating the standard deviations. In both panels, the error bars along the x-axis indicate 68 percentile ranges.         
} 
\label{fig:empirical}    
\end{figure*} 

Simple, idealized models of \lya\ radiation transfer with a central ionizing point source surrounded by a homogeneous, spherically symmetric shell of gas with a range of neutral hydrogen column density ($N(\HI)$), dust opacity, velocity, and temperature \citep[the so-called ``shell model'', see e.g.,][]{Zheng02,Verhamme06} have been surprisingly successful in explaining a large variety of \lya\ line profiles \citep[e.g.,][]{Hashimoto15,Gronke17}. Using the shell model, \citet{Verhamme18} found a correlation between FWHM and $V_{\rm offset}$ \citep[see also][]{Claeyssens19}. The left panel of Fig.~\ref{fig:empirical} shows the weighted average \voff, measured from the median stacked CGM absorption (\HI\ and \CIV) profiles, against the median FWHM of the low-- and high--FWHM$_{\rm resolved}$ sub-samples (blue squares). Consistent with the model prediction, the stacked CGM absorption profiles show larger velocity offsets for the high--FWHM$_{\rm resolved}$ sub-sample (Table~\ref{tab:offset_summary}). The dashed line shows the empirical relation between \voff\ and FWHM obtained by \citet[]{Verhamme18} from the sample of LAEs with known systemic redshifts  as indicated by the star symbols. \citet{Verhamme18} used the {\sc lst\_linefit} routine of \citet{Cappellari13}, which includes a procedure for the rejection of outliers, and obtained a slope of $0.9\pm0.14$, an intercept of $-34\pm60$~\kms, and an intrinsic scatter of $72\pm12$~\kms. The slope (0.89) and intercept ($-$58~\kms) we obtain from the stacked CGM absorption profiles are fully consistent with \citet{Verhamme18}. Note that the trend between \voff\ and FWHM found by \citet{Verhamme18} was determined via observations of the interstellar medium (ISM) properties (nebular emission lines) whereas we confirm the same trend using CGM observations.    

\citet{Steidel10} obtained a mean velocity offset of $445\pm27$~\kms\ between \lya\ and systemic redshifts defined by the $\rm H\alpha$ lines for a sample of 41 $z\approx2.3$ LBGs. The total baryonic masses estimated for those LBGs are $\gtrsim10^{10}-10^{11.5}$~\Msun. Using a sample of 22 NB-selected (with a typical bandwidth of $\approx100$~\AA) LAEs with \lya\ equivalent widths $>50$~\AA, \citet{Shibuya14} obtained an average velocity offset between \lya\ and nebular redshifts of $234\pm9$~\kms. The stellar mass ($M_{\ast}$) estimates for the LAEs in their sample range between $\approx10^{9}$ and $10^{10}$~\Msun. Clearly, the NB-selected LAEs exhibit a smaller velocity offset compared to the broadband-- (UV color) selected LBGs \cite[as already noted by][]{Hashimoto13err,Hashimoto15,Shibuya14}. \citet{Hashimoto15} argued that the low \voff\ of LAEs compared to LBGs are related to smaller $N(\HI)$ in LAEs. Note that, both LBGs and NB-selected LAEs show higher velocity offsets (by factors of $\approx2.6$ and $\approx1.4$, respectively) compared to what we measure for the MUSE-detected LAEs.   

Using the mean \HI\ CGM absorption profile of $\approx$300 UV color selected galaxies in the redshift range $2-3$, \citet{Rakic11} estimated $V_{\rm offset}=295\pm35$~\kms, which is $\approx1.7$ times higher than what we obtained for our sample. The galaxies in \citet{Rakic11} were drawn from \citet{Steidel10} with a typical halo mass of $\sim10^{12}$~\Msun\ \citep[]{Rakic13}. Using clustering properties of LAEs, \citet{Khostovan19} found a strong, redshift-independent correlation between $L$(\lya) normalized by the characteristic line luminosity, $L^{\star}(z)$, and dark matter halo mass. According to their Eq.~13, the median $L$(\lya) of our sample of $\approx10^{42}$~$\rm erg~s^{-1}$ ($L({\rm Ly\alpha})/L^{\star}(z)=0.2$)\footnote{$\log L^{\star}(z=3.3)/\rm erg~s^{-1} = 42.68^{+0.07}_{-0.06}$, see Table~2 of \citet{Khostovan19}} would correspond to a halo mass of $M_{h} \sim10^{10.8}$~\Msun, corresponding to a stellar mass of $M_{\ast}\sim10^{8.0}$~\Msun\ \citep[]{Moster13}, assuming LAEs are average main sequence galaxies. Additionally, the median SFR ($1.3$~\SFR) of our sample corresponds to $M_{*} \sim 10^{8.6}\rm M_{\odot}$ \citep[]{Behroozi19} and $M_{h}\sim 10^{11.1} \rm M_{\odot}$ \citep[]{Moster13}. Clearly, the MUSE-detected LAEs in our sample are, on average, at least an order of magnitude lower in mass than the LBG sample of \citet{Rakic11}. Higher mass galaxies tend to have higher SFR which, in turn, can drive high velocity, galactic-scale winds causing higher (red) shift of the \lya\ emission line.

We find a positive (negative) trend between \voff\ and SFR (EW$_0$), consistent with the findings of \citet{Erb14}. The anti-correlation between \voff\ and EW$_0$ is understood in terms of higher optical depth of gas with near systemic velocity \citep[]{Steidel10,Erb14}. The right panel of Fig.~\ref{fig:empirical} shows the \voff\ measured from the median stacked profiles against the median SFRs of the low-- and high--SFR sub-samples. In addition, we show the \voff\ measurement from \citet{Rakic11} for their sample of $z\approx2.3$ LBGs with a median SFR of $\approx25~\rm M_{\odot}~yr^{-1}$ \citep{Turner14, Steidel14}. A positive trend between \voff\ and SFR is evident in the $\log-\log$ plot. A linear least-squares fit to the data points results in a slope of $0.16\pm0.03$ and an intercept of $2.26\pm0.02$, indicating a sub-linear relationship \voff~$\propto \rm SFR^{0.16\pm0.03}$. The relation holds over almost 2 orders of magnitude range in SFR. We note here that the low--$\log \rm SFR$ bin is dominated by upper limits (24/33). Thus, the inferred slope of the trend will be shallower if the true SFR values are much smaller than the estimated upper limits.

Note that the SFRs in our sample are not corrected for dust, whereas the SFRs for the LBG sample of \cite{Rakic11} are dust-corrected. Using the mean $\beta_{\rm UV}$ of $-2.03$ estimated for $\sim0.1L^{\star}$ galaxies at $z\approx4$ by \citet{Bouwens14} and the relationship between $\beta_{\rm UV}$ and UV extinction ($A_{1600}$) from \citet{Meurer99}\footnote{$A_{1600} = 4.43+1.99\times \beta_{\rm UV}$}, we obtain a mild $\approx0.15$~dex correction in SFR for our sample. Incorporating such a correction factor in SFR provides a consistent best-fitting relationship between \voff\ and SFR (i.e., a slope of $0.17 \pm 0.04$ and an intercept of $2.23 \pm 0.02$).

The open diamonds and the open triangles in the right panel of Fig.~\ref{fig:empirical} represent individual LAEs from \citet{Shibuya14} and \citet{Trainor15}, respectively, for which the SFRs and \voff\ values, measured from non-resonant nebular lines, are known. The SFRs for the \citet{Trainor15} sample are calculated from the $\rm H\alpha$ luminosities using \citet{Kennicutt98} relation. The SFRs of all these LAEs have been corrected to the \citet{Chabrier03} IMF. The median SFRs and the median \voff\ values of these samples are indicated by the corresponding filled symbols. If we include these two points in the fit, we obtain a slope of $0.15\pm0.03$ and an intercept of $2.26\pm0.02$, which are fully consistent with what we obtained earlier. Here standard deviations of the individual measurements are used as uncertainties. The results remain consistent (slope~$=0.12\pm0.03$ and intercept~$=2.25\pm0.02$) within the $1\sigma$ allowed ranges in slope and intercept even if we use standard errors instead of standard deviations.

The correlation between SFR and \voff\ can be explained as follows. Galaxies with higher SFRs are likely to drive higher velocity winds. Higher velocity winds will enhance and shift the red \lya\ peak to a longer wavelength resulting in a larger velocity offset \citep[see Fig.~8 of][for example]{Laursen09}. Using the scaling relations between SFR and $M_{\ast}$ \citep[SFR~$\propto M_{\ast}$ at $z \approx 4$; see e.g., Fig.~3 of][]{Behroozi19}, and between $M_{\ast}$ and $M_{h}$ \citep[$M_{\ast}\propto M_{h}^{2}$ at $z \approx 4$; e.g.,][]{Moster13}, we obtain \voff~$\propto V_{\rm cir}$, where $V_{\rm cir}~(\propto M_{h}^{1/3} \propto M_{\ast}^{1/6} \propto \rm SFR^{1/6})$ is the halo circular velocity. It is interesting to note that in models of momentum driven galactic outflows the wind speed scales as $V_{\rm cir}$ \citep[e.g.,][]{Murray05,Heckman15}. Moreover, models of \lya\ radiative transfer suggest that \voff\ is twice the shell expansion velocity \citep[e.g.,][]{Verhamme06}. Indeed, \cite{Rakic11} found that for $z\approx 2.3$ LBGs, \voff\ is about twice the blueshift of the interstellar absorption lines thought to arise in galactic winds. Hence, if the \lya\ emission is back scattered off an outflowing medium, we expect \voff~$\propto V_{\rm cir}$ which is consistent with our results. Alternatively, a static medium (or a medium without a clear bulk flow) with higher $N(\HI)$ for higher $V_{\rm cir}$ can also explain the correlation.

\section{Summary \& Conclusions}      
\label{sec:con}

Determining accurate redshifts for LAEs is challenging owing to the resonant scattering of \lya\ photons with neutral hydrogen present in the ISM and in the CGM. Here we use CGM absorption lines, detected in the spectra of 8 background quasars, of 96 LAEs at $z\approx3.3$ to calibrate the \lya\ redshifts statistically. These LAEs are detected in 8 MUSE fields centered on the 8 bright quasars with redshifts 3.7--3.8. Our method for calibrating \lya\ redshifts, which was pioneered by \citet{Rakic11}, relies on the assumption that the average (stacked) CGM absorption profiles of LAEs must be centered on the systemic velocity. This simply follows from the fact that the LAEs are randomly oriented with respect to the background quasars. Therefore, the CGM absorption, originating in outflows/accretion/co-rotating gas-disks, should have no preferred line of sight velocities. We measured \voff~$=$~171$\pm$8~\kms\ and 177$\pm$9~\kms, from the median and mean stacked absorption profiles, respectively. The \voff\ obtained for the MUSE-detected LAEs in our sample is smaller than that measured for LBGs in the literature, likely due to the lower masses of LAEs compared to LBGs. \voff\ shows positive trends with FWHM and a negative trend with EW$_0$. Finally, a sub-linear relation is obtained between \voff\ and SFR, which, in turn, suggests that \voff\ scales as the halo circular velocity.

Stacked CGM absorption profiles, as we obtained here, are a powerful tool to calibrate \lya\ redshifts in a statistical manner, which can be applied to samples without systemic redshifts. Nevertheless, obtaining rest-frame optical nebular line diagnostics using future VLT/KMOS, Keck/MOSFIRE, and/or JWST/NIRSpec observations would be extremely useful to determine the systemic redshifts on a galaxy-by-galaxy basis, and to understand the physical properties of these high-$z$, presumably low-mass galaxies.

\vskip0.2cm 
\noindent  
{\it Acknowledgements:} We thank the anonymous referee for useful suggestions. This study is based on observations collected at the European Organisation for Astronomical Research in the Southern Hemisphere under ESO programme(s): 094.A-0131(B), 095.A-0200(A), 096.A-0222(A), 097.A-0089(A), and 099.A-0159(A). 
SM acknowledges support from the Alexander von Humboldt Foundation, Germany. SM thanks Christian Herenz for useful discussion.   
SC gratefully acknowledges support from Swiss National Science Foundation grant PP00P2\_163824. 
JB acknowledges support by FCT/MCTES through national funds by grant UID/FIS/04434/2019 and through Investigador FCT Contract No. IF/01654/2014/CP1215/CT0003. 
NB and JZ acknowledge support from ANR grant ANR-17-CE31-0017 (3DGasFlows). 
AC and JR acknowledge support from the ERC starting grant 336736-CALENDS. 
MA acknowledges support from European Union's Horizon 2020 research and innovation programme under Marie Sklodowska-Curie grant agreement No 721463 to the SUNDIAL ITN, and from the Spanish Ministry of Economy and Competitiveness (MINECO) under grant number AYA2016-76219-P. MA also acknowledges support from the Fundaci\'on BBVA under its 2017 programme of assistance to scientific research groups, for the project ``Using machine-learning techniques to drag galaxies from the noise in deep imaging''.  
FL and AV acknowledge support from the ERC starting grant ERC-757258-TRIPLE.

\appendix 
\section{Stacked profiles generated using the original quasar spectra}  

\begin{figure*}
\centerline{\vbox{
\centerline{\hbox{ 
\includegraphics[trim=00 140 00 230,clip,width=0.7\textwidth,angle=00]{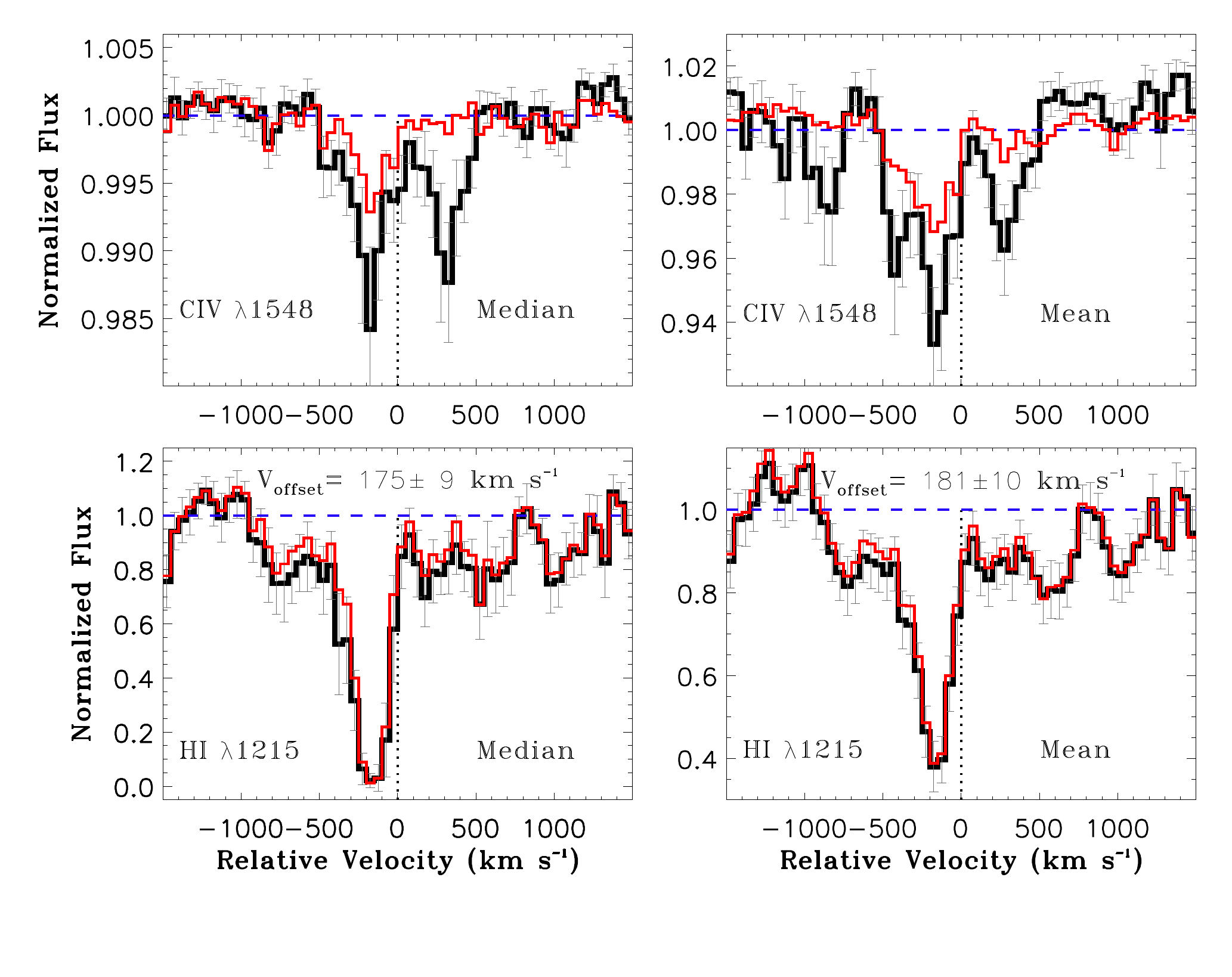} 
}}
}}   
\vskip-1.0cm  
\caption{Median (left) and mean (right) stacked CGM absorption profiles of \HI\ (bottom) and \CIV\ (top) generated using the original quasar spectra. The corresponding profiles using {\tt PODPy} recovered spectra are shown in red for comparison. The strong \HI\ profiles do not show a considerable difference. The relatively weak \CIV\ transition look noisier, particularly for the mean stack, compared to the {\tt PODPy} recovered stack. The weaker member of the \CIV~$\lambda\lambda1548,1550$ doublet, which is eliminated by {\tt PODPy}, is also seen here. The \voff\ values measured from the stacked \HI\ profiles, as indicated in the figure, are fully consistent with those obtained in Fig.~\ref{fig:profiles}. Thus, the use of {\tt PODPy} recovered spectra instead of the original quasar spectra does not have a significant effect on the \voff\ measurements.            
}   
\label{fig:prof_comp}      
\end{figure*}

In Fig.~\ref{fig:prof_comp} we compare the stacked profiles generated using the original spectra and the {\tt PODPy} recovered spectra. From the comparison we conclude that the masking of inconsistent pixels using {\tt PODPy} does not have an appreciable effect on the measured \voff.

%

\def\aj{AJ}%
\def\actaa{Acta Astron.}%
\def\araa{ARA\&A}%
\def\apj{ApJ}%
\def\apjl{ApJ}%
\def\apjs{ApJS}%
\def\ao{Appl.~Opt.}%
\def\apss{Ap\&SS}%
\def\aap{A\&A}%
\def\aapr{A\&A~Rev.}%
\def\aaps{A\&AS}%
\def\azh{AZh}%
\def\baas{BAAS}%
\def\bac{Bull. astr. Inst. Czechosl.}%
\def\caa{Chinese Astron. Astrophys.}%
\def\cjaa{Chinese J. Astron. Astrophys.}%
\def\icarus{Icarus}%
\def\jcap{J. Cosmology Astropart. Phys.}%
\def\jrasc{JRASC}%
\def\mnras{MNRAS}%
\def\memras{MmRAS}%
\def\na{New A}%
\def\nar{New A Rev.}%
\def\pasa{PASA}%
\def\pra{Phys.~Rev.~A}%
\def\prb{Phys.~Rev.~B}%
\def\prc{Phys.~Rev.~C}%
\def\prd{Phys.~Rev.~D}%
\def\pre{Phys.~Rev.~E}%
\def\prl{Phys.~Rev.~Lett.}%
\def\pasp{PASP}%
\def\pasj{PASJ}%
\def\qjras{QJRAS}%
\def\rmxaa{Rev. Mexicana Astron. Astrofis.}%
\def\skytel{S\&T}%
\def\solphys{Sol.~Phys.}%
\def\sovast{Soviet~Ast.}%
\def\ssr{Space~Sci.~Rev.}%
\def\zap{ZAp}%
\def\nat{Nature}%
\def\iaucirc{IAU~Circ.}%
\def\aplett{Astrophys.~Lett.}%
\def\apspr{Astrophys.~Space~Phys.~Res.}%
\def\bain{Bull.~Astron.~Inst.~Netherlands}%
\def\fcp{Fund.~Cosmic~Phys.}%
\def\gca{Geochim.~Cosmochim.~Acta}%
\def\grl{Geophys.~Res.~Lett.}%
\def\jcp{J.~Chem.~Phys.}%
\def\jgr{J.~Geophys.~Res.}%
\def\jqsrt{J.~Quant.~Spec.~Radiat.~Transf.}%
\def\memsai{Mem.~Soc.~Astron.~Italiana}%
\def\nphysa{Nucl.~Phys.~A}%
\def\physrep{Phys.~Rep.}%
\def\physscr{Phys.~Scr}%
\def\planss{Planet.~Space~Sci.}%
\def\procspie{Proc.~SPIE}%
\let\astap=\aap
\let\apjlett=\apjl
\let\apjsupp=\apjs
\let\applopt=\ao
\bibliographystyle{mn}
\bibliography{mybib_new.bib}
\bsp 

\label{lastpage} 
\label{lastpage} 
\end{document}

%% file: table.tex
\begin{table*}  
\begin{threeparttable}[b] 
\caption{Velocity offset measurements for different sub-samples}        
\begin{tabular}{p{3.0cm}p{0.9cm}p{0.8cm}rrp{0.5cm}p{1.0cm}p{0.9cm}p{0.9cm}p{0.9cm}p{0.9cm}c}   
\hline 
Sub-sample                      &  Threshold &  Median   &   16$^{\rm th}$  &   84$^{\rm th}$ &  N$_{\rm LAE}$ &  \voff\       & \voff\       &   \voff\        &   \voff\        &  \voff            &  \voff\        \\ 
                                &            &           &   percentile     &   percentile    &                &   (\HI)       &  (\HI)       &   (\CIV)        &   (\CIV)        &  (\HI+\CIV)       &  (\HI+\CIV)    \\ 
                                &            &           &                  &                 &                &   (Median)    &  (Mean)      &   (Median)      &   (Mean)        &  (Median)         &  (Mean)        \\ 
(1)                             & (2)        & (3)       & (4)              & (5)             & (6)            & (7)           & (8)          & (9)             & (10)            & (11)              & (12)           \\ 
\hline  
 Low--$\rho$                    		&  164.5     &  114.6    &   66.1           &   148.0         &    48     &  180$\pm$11 & 184$\pm$12 & 157$\pm$23 &  194$\pm$23 & 175$\pm$ 9 & 186$\pm$10     \\ 
High--$\rho$                    		&  164.5     &  214.2    &   183.7          &   251.2         &    48     &  158$\pm$15 & 158$\pm$15 & 196$\pm$23 &  181$\pm$25 & 169$\pm$12 & 164$\pm$12    \\ \\       
 Low--\zlae\                   			&  3.335     &  3.083    &   2.959          &   3.267         &    48     &  189$\pm$14 & 183$\pm$12 & 187$\pm$18 &  191$\pm$15 & 188$\pm$11 & 186$\pm$ 9    \\ 
High--\zlae\                    		&  3.335     &  3.556    &   3.400          &   3.655         &    48     &  160$\pm$ 8 & 155$\pm$10 & 146$\pm$40 &  196$\pm$31 & 159$\pm$ 7 & 158$\pm$ 9    \\ \\ 
 Low--FWHM                      		&  239.7     &  195.0    &   161.6          &   227.2         &    48     &  150$\pm$14 & 161$\pm$15 & 133$\pm$20 &  172$\pm$24 & 144$\pm$11 & 164$\pm$12    \\ 
High--FWHM                      		&  239.7     &  289.2    &   255.5          &   356.6         &    48     &  190$\pm$12 & 184$\pm$12 & 211$\pm$20 &  235$\pm$24 & 195$\pm$10 & 194$\pm$10    \\ \\ 
 Low--FWHM$_{\rm resolved}^{a}$			&  253.1     &  214.2    &   184.2  	    &   237.3         &    43     &  141$\pm$13 & 148$\pm$14 & 121$\pm$17 &  148$\pm$23 & 133$\pm$10 & 148$\pm$11    \\ 
High--FWHM$_{\rm resolved}^{a}$			&  253.1     &  290.5    &   258.4          &   376.2         &    43     &  194$\pm$15 & 191$\pm$15 & 217$\pm$22 &  226$\pm$21 & 201$\pm$12 & 202$\pm$12    \\ \\ 
 Low--FWHM$_{{\rm matched}-z_{\rm peak}}^{b}$	&  239.7     &  201.9    &   164.3          &   228.0         &    31     &  139$\pm$14 & 147$\pm$17 & 116$\pm$24 &  133$\pm$27 & 133$\pm$12 & 143$\pm$14    \\ 
High--FWHM$_{{\rm matched}-z_{\rm peak}}^{b}$	&  239.7     &  290.5    &   252.4          &   387.0         &    31     &  199$\pm$12 & 196$\pm$13 & 211$\pm$34 &  233$\pm$20 & 200$\pm$11 & 206$\pm$10    \\ \\ 
 Low--$z_{\rm matched-FWHM}^{c}$   		&  3.335     &  3.083    &   3.000          &   3.305         &    37     &  180$\pm$18 & 172$\pm$11 & 183$\pm$21 &  184$\pm$17 & 181$\pm$13 & 175$\pm$ 9    \\ 
High--$z_{\rm matched-FWHM}^{c}$   		&  3.335     &  3.570    &   3.400          &   3.660         &    37     &  157$\pm$ 7 & 166$\pm$10 & 139$\pm$71 &  122$\pm$53 & 156$\pm$ 6 & 164$\pm$ 9     \\ \\ 
 Low--$\log L$(\lya)          			&  41.97     &  41.76    &   41.52          &   41.91         &    48     &  161$\pm$11 & 160$\pm$14 & 153$\pm$29 &  205$\pm$23 & 159$\pm$10 & 172$\pm$11    \\ 
High--$\log L$(\lya)            		&  41.97     &  42.28    &   42.05          &   42.52         &    48     &  185$\pm$14 & 186$\pm$13 & 219$\pm$46 &  252$\pm$76 & 187$\pm$13 & 187$\pm$12    \\  \\ 
 Low--$\log {\rm SFR}^{d}$            		&  $-0.18$   &  $-0.38$  &   $-0.50$        &   $-0.26$       &    33     &  158$\pm$10 & 161$\pm$12 & 147$\pm$22 &  135$\pm$15 & 156$\pm$ 9 & 150$\pm$ 9    \\ 
High--$\log {\rm SFR}^{d}$            		&  $-0.18$   &  0.23     &   0.01           &   0.49          &    30     &  204$\pm$15 & 192$\pm$13 & 210$\pm$40 &  173$\pm$34 & 204$\pm$14 & 189$\pm$12    \\ \\ 
 Low--EW$_{0}^{d}$                  		&  63.3      &  33.2     &   19.8           &   54.1          &    29     &  203$\pm$11 & 199$\pm$12 & 220$\pm$37 &  171$\pm$24 & 204$\pm$10 & 193$\pm$10    \\ 
High--EW$_{0}^{d}$                  		&  63.3      &  94.0     &   68.8           &   126.0         &    30     &  143$\pm$ 9 & 146$\pm$11 & 144$\pm$17 &  117$\pm$14 & 143$\pm$ 7 & 134$\pm$ 8     \\ 
\hline 
\end{tabular} 
\label{tab:offset_summary}    
Notes-- (1) The sub-sample for which \voff\ is measured. 
(2) The threshold value of the parameter based on which the sub-sample is made:  
$\rho$ in kpc, FWHM in \kms, $L$(\lya) in $\rm erg~s^{-1}$, SFR in $\rm M_{\odot}~yr^{-1}$, and $\rm EW_{0}$ in \AA. 
(3) The median value of the parameter for the sub-sample. 
(4) 16$^{\rm th}$ percentile of the parameter. 
(5) 84$^{\rm th}$ percentile of the parameter. 
(6) Number of LAEs in the sub-sample.  
(7) The velocity offset in \kms\ measured from the median stacked \HI\ profile. 
(8) The same as (7) but for the mean stacked \HI\ profile.  
(9) The same as (7) but for the median stacked \CIV\ profile.     
(10) The same as (7) but for the mean stacked \CIV\ profile.     
(11) The weighted average of \voff\ measured from the median stacked \HI\ and \CIV\ profiles.  
(12) The same as (11) but measured from the mean stacked \HI\ and \CIV\ profiles.   \\  
$^{a}$The LAEs with FWHM smaller than the MUSE resolution ($<166$~\kms) are excluded.  \\ 
$^{b}$The \zlae\ is matched for these two sub-samples. \\ 
$^{c}$The FWHM is matched for these two sub-samples. \\  
$^{d}$ The LAEs that are blended with low-$z$ continuum objects are excluded. The threshold value is chosen in such a way that the number of LAEs in both sub-samples are similar. Only the continuum-detected objects above (below) the threshold value are used to construct the high--$\log \rm SFR$ (low--$\rm EW_0$) sub-sample. All the objects below (above) the threshold value are used to construct the low--$\log \rm SFR$ (high--$\rm EW_0$) sub-sample. Only 9/33 and 10/30 LAEs in the low--$\log \rm SFR$ and high--$\rm EW_0$ sub-samples, respectively, are detected in UV continuum. Upper (lower) limits in the low--$\log \rm SFR$ (high--$\rm EW_0$) sub-samples are considered as detections for the median and percentile calculations. 
\end{threeparttable}  
\end{table*}